\documentclass[a4paper,11pt]{article}

\usepackage{jheppub}
\usepackage[T1]{fontenc}
\usepackage{hyperref}
\usepackage{graphicx}
\usepackage{amsmath}
\usepackage{epsfig}
\usepackage{subfig}
\usepackage{xcolor}
\usepackage{natbib}

\def\be{\begin{equation}}
\def\ee{\end{equation}}
\def\bea{\begin{eqnarray}}
\def\eea{\end{eqnarray}}
\def\NO{\nonumber}
\def\gev{\mathrm{~GeV}}

\def\md{\mathrm{d}}

\title{Double longitudinal-spin asymmetries in $J/\psi$ production at RHIC}

\author[a]{Yu Feng}
\author[a]{Hong-Fei Zhang}
\affiliation[a]{Department of Physics, College of Basic Medical Sciences, Army Medical University, Chongqing 400038, China}
\emailAdd{yfeng@ihep.ac.cn}
\emailAdd{hfzhang@ihep.ac.cn}

\abstract{
The double longitudinal-spin asymmetry, $A_{LL}$,
of the $J/\psi$ production in polarized proton-proton collisions is presented in this paper at QCD next-to-leading order.
It is found that the obtained values of $A_{LL}$ are in general consistent with the PHENIX measurements.
Various sets of the long-distance matrix elements (LDMEs) are employed in our calculation to study the possible theoretical uncertainties.
It is found that, for $p_t<5\gev$, all these LDMEs lead to almost the same results,
which are within the tolerance of the experimental data uncertainties.
}

\begin{document}

\maketitle

\bibliographystyle{JHEP}

\section{Introduction}\label{sec:intro}

The spin structure of the proton is one of the most challenging open puzzles in high energy physics.
Deep-inelastic scattering experiments suggest that only 30\% of the proton spin is carried by its constituent quarks,
which challenged our understanding of the internal structure of the proton and inspired a lot of efforts from both experimental and theoretical aspects.
On the experiment side, many programs are dedicated to the precise study of the proton spin structure.
On the theory side, several frameworks have been proposed to describe the proton spin as the sum of
quarks and gluons spin contributions~\cite{Jaffe:1989jz,Ji:1996ek,Hatta:2011zs,Chen:2011gn,Wakamatsu:2012ve,Ji:2013fga}.

In the infinite momentum frame, all the contributions to the proton spin can be classified according to the Manohar-Jaffe sum rule~\cite{Jaffe:1989jz},
\begin{equation}
S_p=\frac{1}{2}=\frac{1}{2}\Delta\Sigma+\Delta G+L_q+L_g,
\end{equation}
where $\frac{1}{2}\Delta\Sigma$, $\Delta G$, and $L_{q,g}$ represent the contributions from the quark and antiquark spin,
the gluon helicity, and the orbital angular momentum of the quarks and gluons, respectively.
The quark and gluon helicity distributions can be probed in high-energy scattering processes with the polarized nucleons,
allowing access to $\Delta\Sigma$ and $\Delta G$.

The polarized parton distribution functions have been studied extensively at the CERN,
DESY, JLab, RHIC, and SLAC laboratories for decades (as a review, ses~\cite{Aidala:2012mv}).
Based on these experimental data, the global QCD next-to-leading-order (NLO) analyses of polarized parton distribution functions
(PDF)~\cite{deFlorian:2008mr,Hirai:2008aj,deFlorian:2009vb,Leader:2010rb,Blumlein:2010rn,Nocera:2014gqa} revealed
that only about 30$\%$ of the proton's spin is carried by the quark polarization.
The remaining spin must come from the contributions from gluon polarization and from the orbital angular momentum of quarks and gluons,
among which, those from the gluon polarization are essential in understanding the proton spin puzzle.
A newly theoretical calculation by lattice QCD~\cite{Yang:2016plb} suggests that the gluon spin takes as much as 50$\%$ of the proton's spin.

The best probes into the gluon polarization in nucleon are offered by polarized proton-proton collisions available at RHIC~\cite{Aschenauer:2013woa}.
The measurements of the longitudinal double-spin asymmetries for inclusive jet~\cite{Abelev:2006uq,Abelev:2007vt,Adamczyk:2012qj,Adamczyk:2014ozi}
and $\pi^{0}$~\cite{Adare:2008aa,Adare:2008qb,Adare:2014hsq} obtained at RHIC play an important role in constraining the distribution of the gluon polarization.
The latest global fits~\cite{deFlorian:2014yva,Nocera:2014gqa} that incorporate the inclusive jet~\cite{Adamczyk:2014ozi} and $\pi^0$~\cite{Adare:2014hsq}
double longitudinal-spin asymmetries now find compelling evidence for positive gluon polarization of roughly 0.2 over the range $x>0.05$.
The polarized PDF for gluons in the region $0.01 < x < 0.05$ has been explored with the measurements of $A^{\pi^0}_{LL}$
at midrapidity by STAR~\cite{Abelev:2009pb} and PHENIX~\cite{Adare:2015ozj} collaborations.
Recently, the STAR collaboration published their results on the $A_{LL}$ for dijet
which provides new constrains on $\Delta g(x)$ in the region $x\sim 0.01$.
Their measurements on $A^{\pi^0}_{LL}$~\cite{Adam:2018cto} at forward rapidities
can even help to constrain $\Delta g(x)$ down to $x\sim 10^{-3}$.

Heavy quarkonium also provides a useful laboratory to access the polarized gluon distribution.
At RHIC energies, heavy quarkonium is dominated by gluon-gluon scattering,
therefore, the corresponding double longitudinal-spin asymmetries $A_{LL}$
are expected to be sensitive to the polarized gluon distribution in nucleon.
On the other hand, heavy quarkonium can be calculated perturbatively,
exploiting the nonrelativistic QCD (NRQCD) factorization formalism~\cite{Bodwin:1994jh},
which allows one to organize the theoretical calculations as double expansions in the QCD coupling constant $\alpha_s$ and the heavy quark relative velocity $v$.
Great phenomenological progress has also been made to test the quarkonium production mechanisms in the past decades~\cite{Brambilla:2014jmp}.
Particularly, the $J/\psi$ hadroproduction data are described remarkably well by NRQCD at QCD NLO
\cite{Gong:2008ft,Butenschoen:2010rq,Ma:2010yw,Butenschoen:2011yh,Ma:2010jj,Bodwin:2014gia}.
On the other hand, the double longitudinal-spin asymmetry in $J/\psi$ production has been studied
at QCD leading order (LO) within both color-singlet~\cite{Morii:1993ja,Doncheski:1994ff}
and color-octet mechanisms~\cite{Teryaev:1996sr,Gupta:1996dt,Klasen:2003zn} in the NRQCD framework.
However, the relevant calculation at QCD NLO is still lacking.
Recently, the measurements~\cite{Adare:2016cqe} of $A^{J/\psi}_{LL}$ by PHENIX collaboration came out,
which makes it possible to study the proton spin in the $J/\psi$ production processes.
Since the QCD corrections to the $J/\psi$ hadroproduction are exceptionally significant,
we, in this paper, study the double longitudinal-spin asymmetry in $J/\psi$ production at QCD NLO.
Using the RHIC data, we will find out whether the polarized gluon PDFs are consistent with the new measurements.

The rest of this paper is organized as follows.
In section~\ref{char:2}, we outline the formalism of our calculation.
In section~\ref{char:3}, we present our numerical results and discuss their phenomenological implications.
Our conclusions are summarized in section~\ref{char:4}.

\section{Calculation of Double Spin Asymmetry}\label{char:2}

\subsection{Double longitudinal-spin asymmetry}
The double longitudinal-spin asymmetry $A_{LL}$ in the polarized proton-proton collisions is defined as
\begin{equation}\label{eq:All}
A_{LL}=\frac{\sigma^{++}-\sigma^{+-}}{\sigma^{++}+\sigma^{+-}}=\frac{\Delta \sigma}{\sigma},
\end{equation}
where $\xi_1$ and $\xi_2$ in $\sigma^{\xi_1\xi_2}$ denote the sign of the helicity of the left- and right-hand-side colliding protons, respectively.
The polarized and unpolarized cross sections, $\Delta\sigma$ and $\sigma$, are defined as
\begin{subequations}\label{eq:sig}
\begin{align}
\label{eq:sig:pol}
\Delta \sigma & = \frac{1}{4}\sum_{\xi_A, \xi_B=\pm}(-1)^{\delta_{\xi_A\xi_B}}\sigma^{\xi_A\xi_B},
\\
\label{eq:sig:unpol}
\sigma & = \frac{1}{4}\sum_{\xi_A, \xi_B=\pm}\sigma^{\xi_A\xi_B}.
\end{align}
\end{subequations}

Exploiting the polarized PDFs, one can rewrite the polarized cross sections in terms of the parton-level polarized cross sections.
Defining $f_{a/A}^+$ ($f_{a/A}^-$) as the PDF of a polarized parton, $a$, in a nucleon, $A$,
with the same polarization as (the opposite polarization to) $a$,
$\sigma^{\xi_A\xi_B}$ can be expressed as
\bea
&&\sigma^{\xi_A\xi_B}=\sum_{a,b}f^{+}_{a/A}\otimes f^{+}_{b/B}\otimes\sigma_{ab}^{\xi_A,\xi_B}
+\sum_{a,b}f^{+}_{a/A}\otimes f^{-}_{b/B}\otimes\sigma_{ab}^{\xi_A,-\xi_B} \NO \\
&&~~~~+\sum_{a,b}f^{-}_{a/A}\otimes f^{+}_{b/B}\otimes\sigma_{ab}^{-\xi_A,\xi_B}
+\sum_{a,b}f^{-}_{a/A}\otimes f^{-}_{b/B}\otimes\sigma_{ab}^{-\xi_A,-\xi_B},
\eea
where $\sigma_{ab}^{\xi_A,\xi_B}$ are the corresponding parton-level polarized cross sections,
and the summations run over all the possible species of the initial partons.
Here, we use $\otimes$ to imply that the parton-level cross sections should convolute with the PDFs.
With the above definitions, we can express the polarized cross sections in a more explicit form as
\bea
&&\Delta\sigma=\sum_{a,b}f^{+}_{a/A}\otimes f^{+}_{b/B}\otimes\frac{1}{4}\sum_{\xi_A,\xi_B=\pm}(-1)^{\delta_{\xi_A\xi_B}}\sigma_{ab}^{\xi_A,\xi_B} \NO \\
&&~~~~-\sum_{a,b}f^{+}_{a/A}\otimes f^{-}_{b/B}\otimes\frac{1}{4}\sum_{\xi_A,\xi_B=\pm}(-1)^{\delta_{\xi_A,-\xi_B}}\sigma_{ab}^{\xi_A,-\xi_B} \NO \\
&&~~~~-\sum_{a,b}f^{-}_{a/A}\otimes f^{+}_{b/B}\otimes\frac{1}{4}\sum_{\xi_A,\xi_B=\pm}(-1)^{\delta_{-\xi_A,\xi_B}}\sigma_{ab}^{-\xi_A,\xi_B} \NO \\
&&~~~~+\sum_{a,b}f^{-}_{a/A}\otimes f^{-}_{b/B}\otimes\frac{1}{4}\sum_{\xi_A,\xi_B=\pm}(-1)^{\delta_{-\xi_A,-\xi_B}}\sigma_{ab}^{-\xi_A,-\xi_B}. \label{eqn:ds}
\eea
If we adopt the following definitions,
\bea
&&\Delta f_{a/A}=f_{a/A}^+-f_{a/A}^-, \NO \\
&&\Delta\sigma_{ab}=\frac{1}{4}\sum_{\xi_a,\xi_b=\pm}(-1)^{\delta_{\xi_a\xi_b}}\sigma_{ab}^{\xi_a,\xi_b},
\label{eqn:sigab:pol}
\eea
we can rewrite Equation (\ref{eqn:ds}) in a more compact form as
\bea
\Delta\sigma=\sum_{a,b}\Delta f_{a/A}\otimes\Delta f_{b/B}\otimes\Delta\sigma_{ab}.
\eea

According to the NRQCD factorization formalism, the cross section for $J/\psi$ hadroproduction are factorized as the
perturbatively calculable short-distance coefficients (SDCs),
which produce on-shell $c\bar{c}$ pairs with definite color and angular-momentum,
and the non-perturbative long-distance matrix elements (LDMEs),
which describe the long-distance processes of the hadronization of these $c\bar{c}$ pairs.
The parton-level polarized cross sections thus can also be factorized within the NRQCD framework.
Explicitly, we have
\bea
&&\Delta\sigma(A+B\rightarrow J/\psi+X) \NO \\
&&~~=\sum_{a,b}\int\md x_1\md x_2\Delta f_{a/A}(x_1)\Delta f_{b/B}(x_2)
\Delta\sigma(a+b\rightarrow J/\psi+X) \NO \\
&&~~=\sum_{a,b,n}\int\md x_1\md x_2\Delta f_{a/A}(x_1)\Delta f_{b/B}(x_2)
\Delta\hat{\sigma}(a+b\rightarrow c\overline{c}(n)+X)\langle{\cal O}^{J/\psi}(n)\rangle, \label{eq:polDsig}
\eea
where the indices $a$, $b$ run over all parton species and $n$ runs over the colors and angular-momenta of the intermediate $c\bar{c}$ states.
Since the polarized PDFs and the LDMEs can be found in many published papers,
the only missing elements in Equation (\ref{eq:polDsig}) are the parton-level polarized SDCs, $\Delta\hat{\sigma}(a+b\rightarrow c\overline{c}(n)+X)$,
the evaluation of which will be addressed in the following subsection.

\subsection{The polarized SDCs at QCD NLO}
The $J/\psi$ meson can be produced directly or via the feed down from higher excited states,
the latter of which accounts about 20\%-30\% of the prompt $J/\psi$ events in RHIC experiments,
and thus is considered not important for the double longitudinal-spin asymmetry.
In this paper, we only take count of the contributions from the directly produced $J/\psi$.
According to the NRQCD factorization, four intermediate states,
$^3S_1^{[1]}$, $^1S_0^{[8]}$, $^3S_1^{[8]}$, and $^3P_J^{[8]}$, are involved in our calculations.

At QCD LO, the color-singlet $J/\psi$ can only be produced via the gluon-gluon fusion, namely
\begin{equation}\label{eq:locs}
g+g\rightarrow c\overline{c}(^3S_1^{[1]})+g,
\end{equation}
while the color-octet $c\bar{c}$ states can be produced in the following three types of processes,
\begin{equation}\label{eq:lo}
\begin{split}
&g+g\rightarrow c\overline{c}(n)+g, \\
&g+q(\overline{q})\rightarrow c\overline{c}(n)+q(\overline{q}), \\
&q+\overline{q}\rightarrow c\overline{c}(n)+g,
\end{split}
\end{equation}
where $n$ = $^1S_0^{[8]}$, $^3S_1^{[8]}$, $^3P_J^{[8]}$,
and $q$, here, denotes light quarks, which, for the $J/\psi$ productions, are $u$, $d$, and $s$.

At QCD NLO, we need to consider both real and virtual correction processes,
the latter of which are comprised of loop and counter-term contributions.
In our calculation, we adopt on-shell renormalization scheme to renormalize the quark and gluon wave functions,
and modified-minimum-subtraction ($\overline{\mathrm{MS}}$) scheme to renormalize the strongly coupling constant.
The renormalization constants are obtained as
\begin{eqnarray}
&&\delta Z^{OS}_{m}=-3C_{F}\frac{\alpha_{s}}{4\pi}\Big[\frac{1}{\epsilon_{UV}}-\gamma_{E}
+\ln\frac{4\pi\mu^{2}_{r}}{m^{2}_{c}}+\frac{4}{3}\Big], \NO \\
&&\delta Z^{OS}_{2}=-C_{F}\frac{\alpha_{s}}{4\pi}\Big[\frac{1}{\epsilon_{UV}}+\frac{2}{\epsilon_{IR}}-3\gamma_{E}
+3\ln\frac{4\pi\mu^{2}_{r}}{m^{2}_{c}}+4\Big], \NO \\
&&\delta Z^{OS}_{2l}=-C_{F}\frac{\alpha_{s}}{4\pi}\Big[\frac{1}{\epsilon_{UV}}-\frac{1}{\epsilon_{IR}}\Big], \NO \\
&&\delta Z^{OS}_{3}=\frac{\alpha_{s}}{4\pi}\Big[(\beta_{0}-2C_{A})\Big(\frac{1}{\epsilon_{UV}}-\frac{1}{\epsilon_{IR}}\Big)\Big], \NO \\
&&\delta Z^{\overline{MS}}_{g}=-\frac{\beta_{0}}{2}\frac{\alpha_{s}}{4\pi}\Big[\frac{1}{\epsilon_{UV}}-\gamma_{E}+\ln(4\pi)\Big],
\end{eqnarray}
where $\mu_r$ is the renormalization scale, $\gamma_E$ is Euler's constant,
$\beta_0=\frac{11}{3}C_A-\frac{4}{3}T_Fn_f$ is the QCD one-loop beta function,
$n_f=3$ is the number of active quark flavors,
and the color factors are given by $T_F=\frac{1}{2}$, $C_F=\frac{4}{3}$, $C_A=3$.
Note that we neglected the contributions of the $c$-quark loop in the gluon self-energy corrections.

QCD correction processes can also be constructed from the LO processes by emitting an additional gluon or splitting a gluon into a quark-antiquark pair,
which are named real-correction processes.
For the CS channel, there are four real-correction processes considered in this paper, namely,
\begin{eqnarray}\label{eq:nlo1}
  &&g+g\rightarrow c\overline{c}(^3S_1^{[1]})+g+g, \NO \\
  &&g+g\rightarrow c\overline{c}(^3S_1^{[1]})+q+\overline{q}, \NO \\
  &&g+q(\overline{q})\rightarrow c\overline{c}(^3S_1^{[1]})+g+q(\overline{q}), \NO \\
  &&q+\overline{q}\rightarrow c\overline{c}(^3S_1^{[1]})+g+g,
\end{eqnarray}
while for each CO state, there are eight such processes~\cite{Gong:2008ft,Gong:2010bk}:
\begin{eqnarray}\label{eq:nlo2}
&g+g\rightarrow c\overline{c}(n)+g+g,~~~
&g+g\rightarrow c\overline{c}(n)+q+\overline{q}, \NO \\
&g+q(\overline{q})\rightarrow c\overline{c}(n)+g+q(\overline{q}),~~~
&q+\overline{q}\rightarrow c\overline{c}(n)+g+g, \NO \\
&q+\overline{q}\rightarrow c\overline{c}(n)+q+\overline{q},~~~
&q+\overline{q}\rightarrow c\overline{c}(n)+q'+\overline{q'}, \NO \\
&q+q\rightarrow c\overline{c}(n)+q+q,~~~
&q+q'\rightarrow c\overline{c}(n)+q+q'.
\end{eqnarray}
where $q$, $q'$ ($\bar{q}$, $\bar{q}'$) denote light quarks (anti-quarks) with different flavors.
Note that we omitted the processes, $g+g\rightarrow c\bar{c}(n)+c+\bar{c}$, in our calculation,
because they are important only in high $p_t$ region, which is not concerned in this paper.

To obtain finite cross sections,
the SDCs for $c\bar{c}(^3P_J^{[8]})$ should be renormalized considering the contributions from the QCD corrections to the $^3S_1^{[8]}$ LDME.
We refer interested readers to References~\cite{Jia:2014jfa, Feng:2017bdu},
where this renormalization procedure are described in detail.

Evaluating the squared amplitudes, we sum over the spins (and colors) of the final-state particles,
while keeping those of the initial-state ones.
Then the initial-spin-and-color-averaged squared amplitudes can be written as
\bea
\mathcal{M}_{\xi_1,\xi_2}\equiv\frac{1}{N_sN_c}\mathcal{A}_{\xi_1,\xi_2}^*\mathcal{A}_{\xi_1,\xi_2},
\eea
where $N_s$ and $N_c$ are the numbers of the initial spin and color states, respectively.
Having this, we can express $\hat{\sigma}_{ab}^{\xi_1\xi_2}$ as
\bea
\mathrm{d}\hat{\sigma}_{ab}^{\xi_1\xi_2}=\frac{1}{2\hat{s}}\mathcal{M}_{\xi_1,\xi_2}\mathrm{d}\phi,
\eea
where $\hat{s}$ is the squared colliding energy of the initial partons,
and $\mathrm{d}\phi$ is the phase-space small element.
Exploiting Equation (\ref{eqn:sigab:pol}), one can obtain the polarized cross sections at the parton level.

\section{Numerical results}\label{char:3}

To numerically evaluate the SDCs, we make use of the FDC program~\cite{Wang:2004du,Wan:2014vka} to generate all the needed FORTRAN source.

In our numerical calculation, we use LHAPDF interface~\cite{Whalley:2005nh} to invoke NNPDFpol1.1~\cite{Nocera:2014gqa} and NNPDF3.0~\cite{Ball:2014uwa},
which are employed in our calculation as the polarized and unpolarized PDFs, respectively.
The colliding energy and the rapidity region are set to be $\sqrt{s}=510\gev$ and $1.2\leq |y|\leq2.2$,
in accordance with the RHIC experiment.
The $c$-quark mass, factorization, renormalization and NRQCD scales are chosen as $m_c=1.5$ GeV,
$\mu_f=\mu_r=\sqrt{4m^2_c+p^2_t}$, and $\mu_{\Lambda}=m_c$, respectively.

Since there are several parallel extractions of the LDMEs, we need to investigate the uncertainties brought about by the different values of them.
Five sets of LDMEs taken from Ref.~\cite{Butenschoen:2011yh,Chao:2012iv,Gong:2012ug, Bodwin:2014gia,Zhang:2014ybe,Sun:2015pia} are collected in Table.~\ref{tab:ldme}
\footnote{Since the CS LDME was not given in Ref.~\cite{Bodwin:2014gia},
we adopt $\langle{\cal O}^{J/\psi}(^{3}S^{[1]}_{1})\rangle=1.16$ GeV as a reasonable choice.}.

\begin{table}[tbp]
  \centering
  \scalebox{0.9}{
  \begin{tabular}{@{\extracolsep{\fill}}*{5}{c}}
  \hline
  Ref.&~$\langle{\cal O}^{J/\psi}(^{3}S^{[1]}_{1})\rangle$~&~$\langle{\cal O}^{J/\psi}(^{1}S^{[8]}_{0})\rangle$~&
  ~$\langle{\cal O}^{J/\psi}(^{3}S^{[8]}_{1})\rangle$~&~$\langle{\cal O}^{J/\psi}(^{3}P^{[8]}_{0})\rangle$~ \\
  ~& (GeV$^3$) & (GeV$^3$)& (GeV$^3$) & (GeV$^5$)~ \\
  \hline
  Butenschon and Kniehl~\cite{Butenschoen:2011yh}&1.32~&3.04 $\times10^{-2}$~& 1.68$\times10^{-3}$& -9.08$\times10^{-3}$ \\
  Chao et~al.~\cite{Chao:2012iv}&1.16~&8.9 $\times10^{-2}$~& 3.0 $\times10^{-3}$&~1.26$\times10^{-2}$ \\
  Gong et~al.~\cite{Gong:2012ug}&1.16~&9.7$\times10^{-2}$&-4.6$\times10^{-3}$&-2.14$\times10^{-2}$ \\
  Bodwin et~al.~\cite{Bodwin:2014gia} & ~ &9.9 $\times10^{-2}$& 1.1 $\times10^{-2}$& 1.1$\times10^{-2}$ \\
  Zhang et~al.~\cite{Zhang:2014ybe,Sun:2015pia} & 0.65 &0.78 $\times10^{-2}$ & 1.08$\times10^{-2}$& 4.52$\times10^{-2}$ \\
  \hline
  \end{tabular}
  }
  \caption{\label{tab:ldme}
  The values of the LDMEs for $J/\psi$ production taken from
  Ref.\cite{Butenschoen:2011yh,Chao:2012iv,Gong:2012ug,Bodwin:2014gia,Zhang:2014ybe,Sun:2015pia}.}
\end{table}

\begin{figure}[!ht]
  \centering
  \includegraphics[width=0.45\textwidth,origin=c]{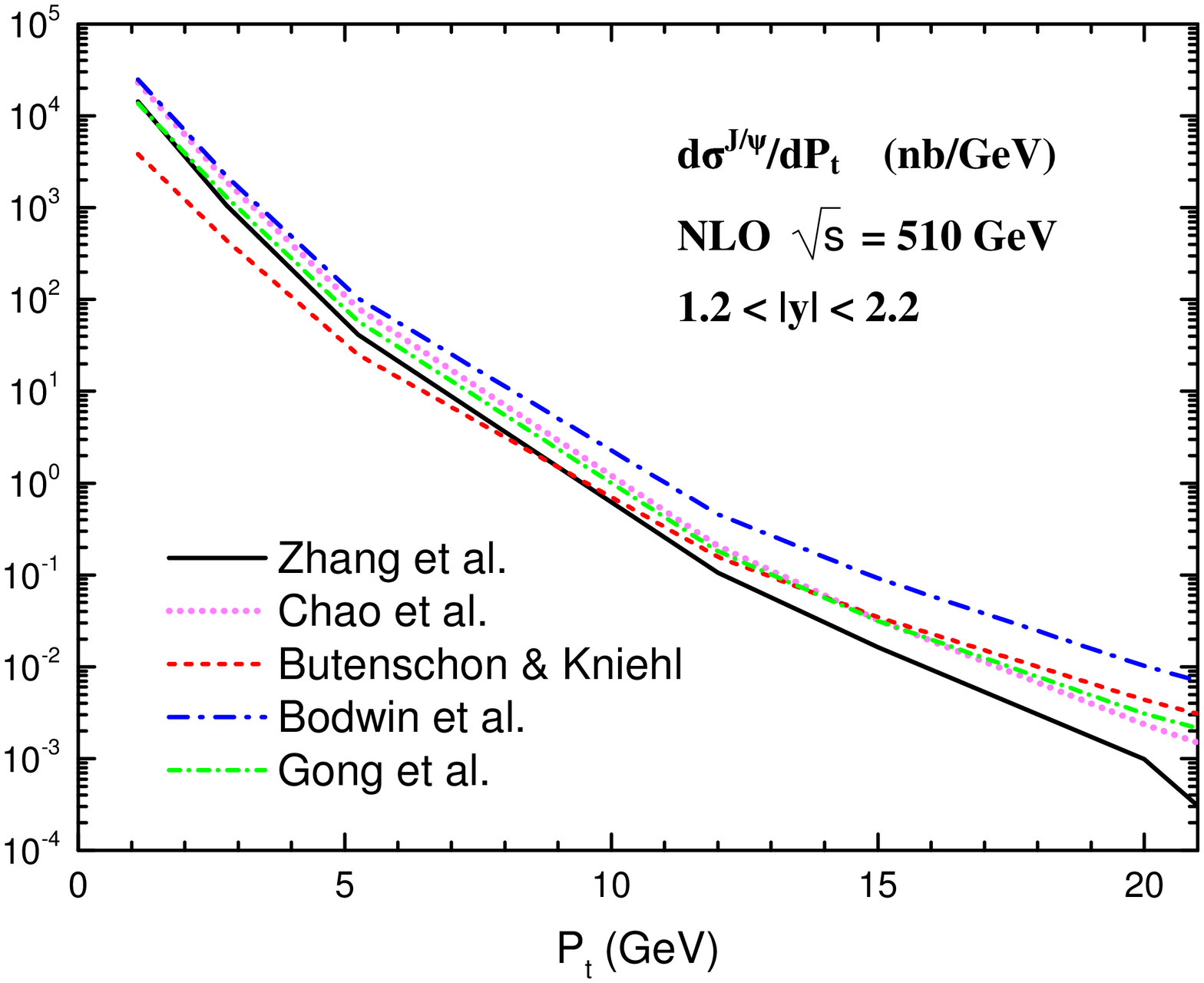}
  \hfill
  \includegraphics[width=0.45\textwidth,origin=c]{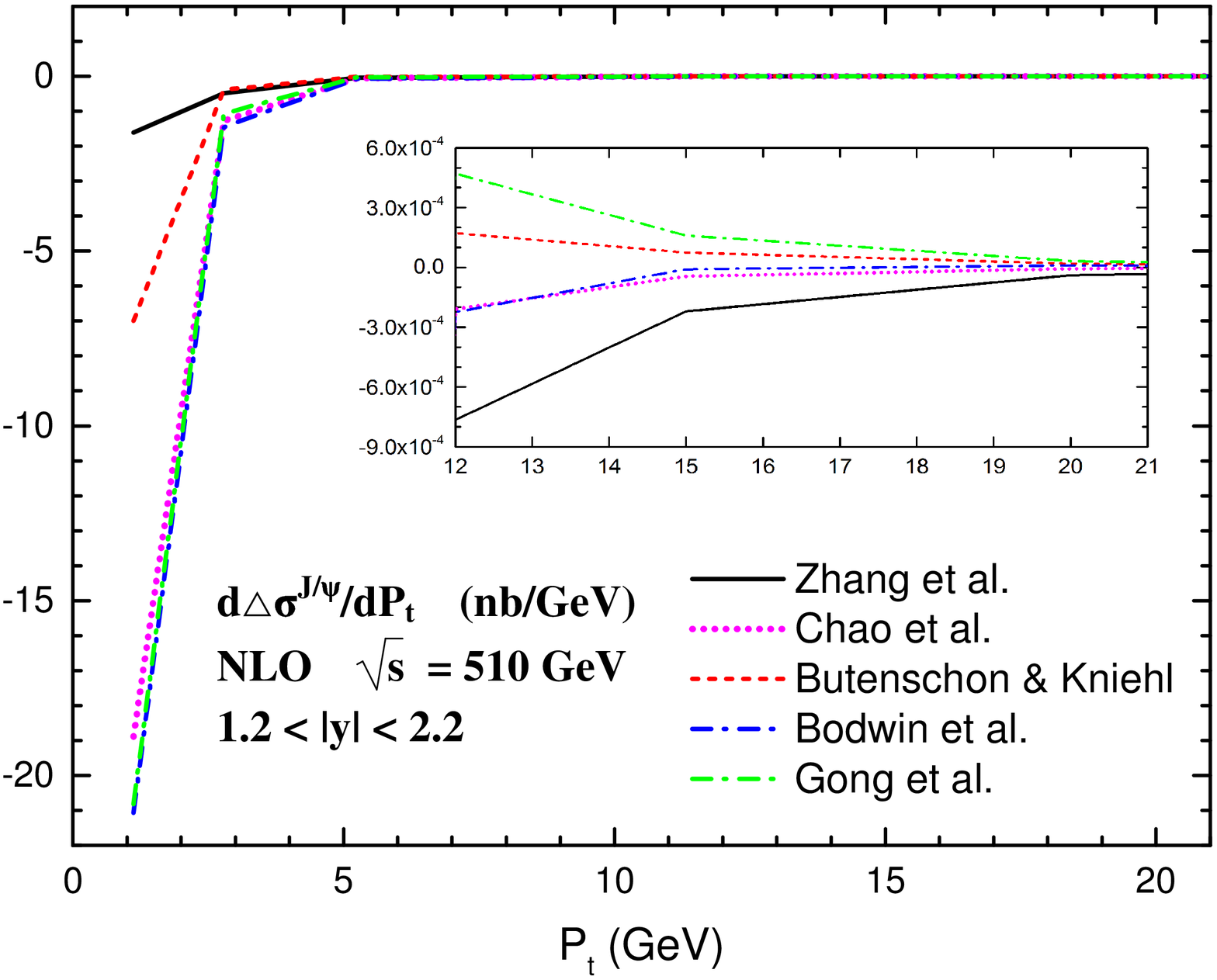}
  \caption{\label{fig:sigpol} The unpolarized (left-hand side) and polarized (right-hand side) cross sections of $J/\psi$ direct production at $\sqrt{s}=510$ GeV up to QCD NLO.}
\end{figure}

In Figure~\ref{fig:sigpol}, we present the unpolarized (L.H.S.) and polarized (R.H.S) differential cross sections with respect to $p_t$.
The curves correspond to different sets of LDMEs.
It should be noted that the results for unpolarized cross sections using different sets of LDMEs do not agree with each other.
Unfortunately, there is no experimental data available in this kinematic region,
thus we cannot judge which one is better to describe the $J/\psi$ yield.
This observation indicate that the $J/\psi$ production at 500 GeV in the region, $1.2<y<2.2$,
can serve as an independent constraint on the LDMEs.
For this reason, we suggest that, PHENIX collaboration do this measurement,
which could contribute enormously to the determination of the NRQCD parameters.
Another interesting observation is that the results for the polarized cross sections employing different sets of LDMEs also differ from each other.
All these results in low $p_t$ regions are negative.
However, some of them change their signs as $p_t$ becomes higher.

\begin{figure}[tbp]
  \centering
  \includegraphics[width=0.8\textwidth]{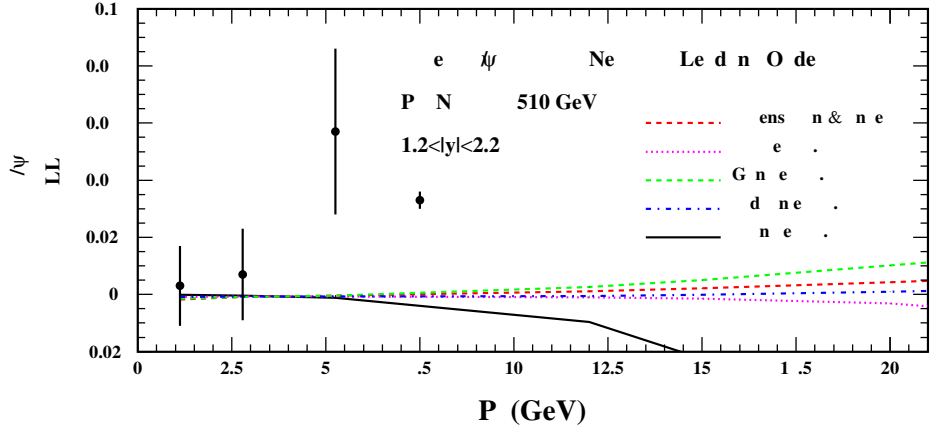}
  \caption{\label{fig:ldme-nlo} The double longitudinal asymmetry $A_{LL}^{J/\psi}$ at QCD NLO with different LDMEs schemes.
  The data points are from Ref.~\cite{Adare:2016cqe}.}
\end{figure}

We are now in a position to discuss the double longitudinal spin asymmetry, $A_{LL}^{J/\psi}$.
Our theoretical results are compared to the experimental data given by PHENIX Collaboration~\cite{Adare:2016cqe} in Figure~\ref{fig:ldme-nlo}.
In low $p_t$ region, say $p_t<5\gev$, all sets of the LDMEs result in the same theoretical prediction that is almost zero.
As $p_t$ increases, these results become distinguishable.
The LDMEs taken from References~\cite{Chao:2012iv, Zhang:2014ybe, Sun:2015pia} lead to negative values,
while the other three sets of LDMEs give positive results.
The largest $p_t$ of the experimental data is about 5 GeV,
below which, the uncertainties brought about by the LDMEs are negligible.
In this sense, we can say that the results obtained from different sets of LDMEs are consistent with each other.
We can see from Figure~\ref{fig:ldme-nlo} that the first two data points are evidently compatible with zero,
while the lower bound of the error bar of the last one is quite above zero.
However, since its uncertainty is huge, it is also considered as compatible with zero~\cite{Adare:2016cqe},
which is consistent with our calculations.

\section{Summary}\label{char:4}

We calculated the QCD corrections to the double longitudinal spin asymmetry of the $J/\psi$ production in polarized proton-proton collisions at RHIC.
To perform a reliable prediction, various sets of NRQCD long-distance matrix elements obtained from different fitting strategies are employed.
The uncertainties brought about by the LDMEs are minor when $p_t$ is smaller than 5 GeV,
and become larger as $p_t$ increases.
Our results are consistent with the RHIC measurements.
To acquire solid information for the $J/\psi$ production and the gluon spin in protons,
we look forward to better precision measurements from the future running of the high-energy hadron colliders.

\acknowledgments

We thank Yan-Qin Ma for helpful discussion.
This work is supported by the National Natural Science Foundation of China (Grants No.11747037).
Y.Feng is also supported by the Army Medical University of PLA of China (No.2016XPY06).

\providecommand{\href}[2]{#2}\begingroup\raggedright\endgroup

\end{document}